\documentclass[12pt]{article}
\usepackage[dvips]{graphicx}
\usepackage{pdproc}

  \makeatletter 
  \def\@cite#1{[#1]} 
  \makeatother    
\begin{document}

\renewcommand{\thefootnote}{\alph{footnote}}

\title{
  Exploring Flavor Structure of SUSY Breaking from B Decays and the
  Unitarity Triangle%
\footnote{
Talk based on Ref.~\cite{Goto:2002xt}, given at the 12th
International Conference on Suypersymmetry and Unification of the
Fundamental Interactions (SUSY04), June 17-23, 2004, in Epochal Tsukuba,
Tsukuba, Japan.
}
}

\author{ TORU GOTO }

\address{ 
YITP, Kyoto University,
Kyoto, 606-8502 Japan
\\ {\rm E-mail: goto@yukawa.kyoto-u.ac.jp}
}

\abstract{
We study effects of SUSY particles in the unitarity triangle analysis
and the CP asymmetries in $b\to s$ decays.
We consider three different SUSY models, the minimal supergravity, SU(5)
SUSY GUT with right-handed neutrinos, and MSSM with U(2) flavor
symmetry, and show that the patterns and correlations of deviations from
the standard model will be useful to discriminate the different SUSY
models in future $B$ experiments.
}

\normalsize\baselineskip=15pt

\section{Introduction}

Goals of $B$ physics are not only to determine the
Cabbibo-Kobayashi-Maskawa (CKM) matrix, but also to search for new
sources of CP violation and flavor mixings.
In the minimal supersymmetric standard model (MSSM), the squark mass
matrices can be such sources.
Since the squark mass matrices depend on the SUSY breaking mechanism,
$B$ physics can provide important information on the origin of the SUSY
breaking.

In this work \cite{Goto:2002xt}, we study the flavor signals in SUSY
models with different flavor structures, namely, the minimal
supergravity (mSUGRA), the SU(5) SUSY GUT with right-handed neutrinos,
and the MSSM with U(2) flavor symmetry \cite{Barbieri:1996ww}.
We consider two cases in the SU(5) SUSY GUT with $\nu_R$.
One is ``degenerate'' case, where all the masses of right-handed
neutrinos are the same, and the other is ``non-degenerate'' case, where
the  $\nu_R$ mass matrix is not proportional to the unit matrix.

We focus on two subjects: the consistency test of the unitarity triangle
and the CP asymmetries in $b\to s$ decays.
We calculate SUSY contributions to the $B_d-\bar{B}_d$,
$B_s-\bar{B}_s$, and $K^0-\bar{K}^0$ mixing amplitudes and show that the
unitarity triangle analysis is useful to discriminate these models.
As for the $b\to s$ decays, we investigate the direct CP asymmetry in
the inclusive $b\to s \gamma$ decay \cite{Kagan:1998bh} and the
mixing-induced CP asymmetry in $B_d\to \phi K_S$ \cite{Barbieri:1997kq}
and $B_d\to K_{\rm CP} \gamma$ \cite{Atwood:1997zr}, where
$K_{\rm CP}$ is a CP eigenstate with a strange quark.
We show that the above models exhibit different pattern of deviations
from the SM.

\section{Unitarity Triangle}

In the SM, the source of the flavor mixing and the CP violation is the
CKM matrix only.
Present knowledge of the parameters in the unitarity triangle
$\bar{\rho}$ and $\bar{\eta}$, which are defined as
$\bar{\rho} + i \bar{\eta}=-\frac{V_{ub}^*V_{ud}}{V_{cb}^*V_{cd}}$,
is obtained by
combining the measured values of $|V_{ub}|$, $\sin2\phi_1$ from the CP
asymmetry in $B_d\to J/\psi K_S$, $\varepsilon_K$, $\Delta m_{B_d}$ and
the lower bound of $\Delta m_{B_s}$.
In the future, we expect that $\Delta m_{B_s}$ and $\phi_3$ are measured
precisely.
At this stage, some mismatch among the observables might be found, which
means an existence of new source of the flavor mixing and/or CP
violation.
We can obtain information about the flavor structure of the SUSY model
from the pattern of the mismatch.

\begin{figure}[htb]
\centering
\includegraphics[scale=0.22,clip]{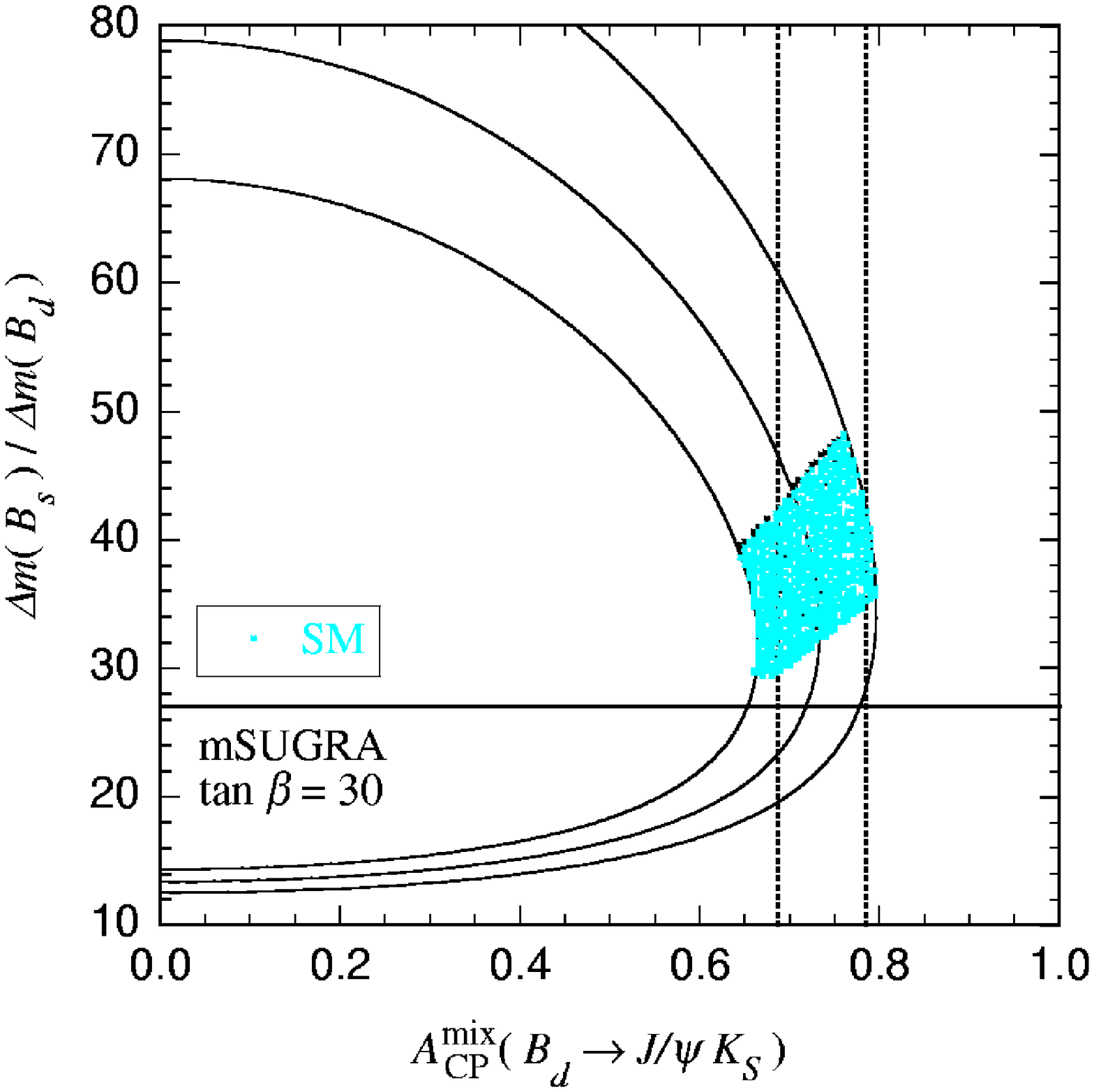}
\includegraphics[scale=0.22,clip]{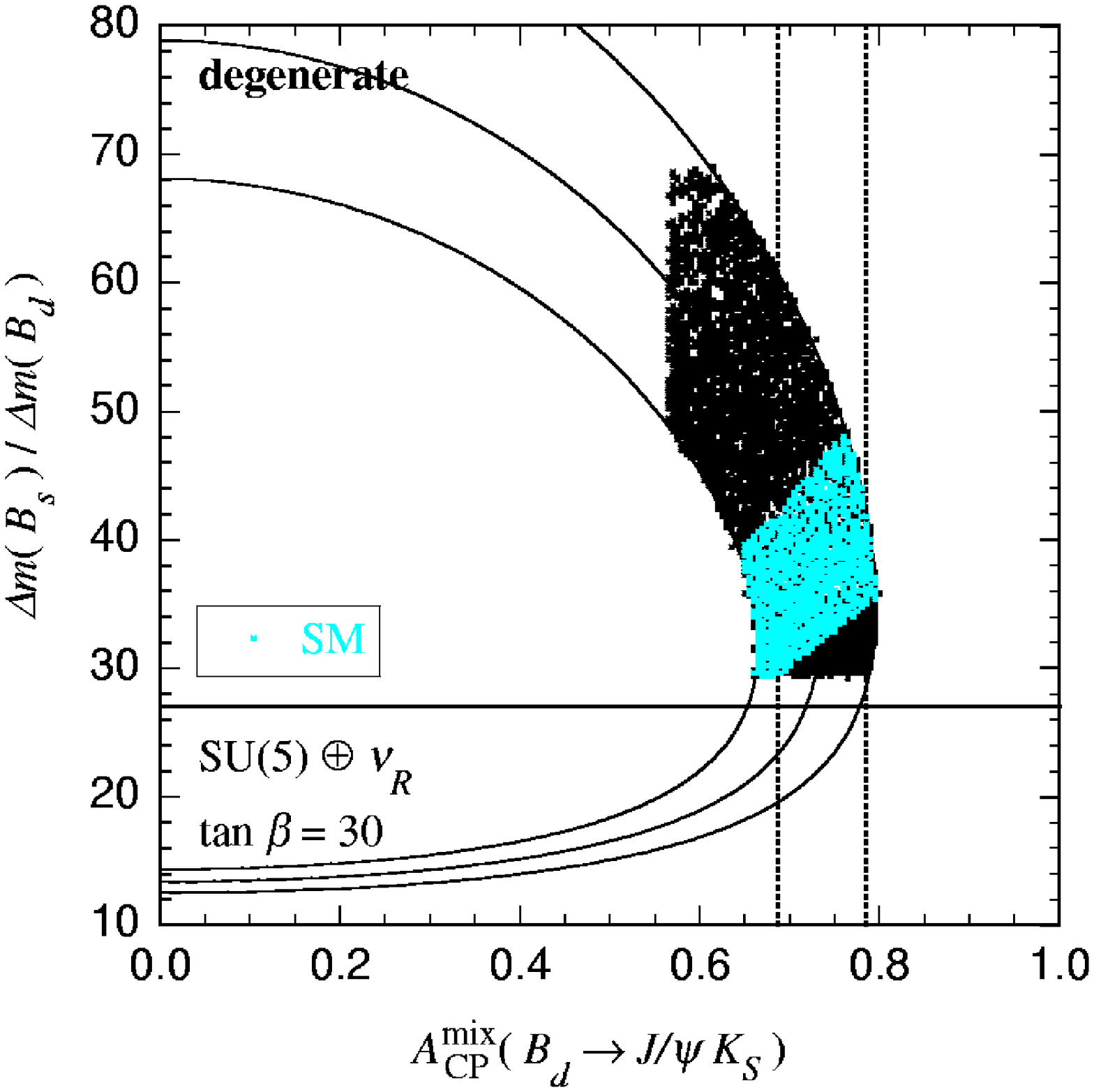}
\includegraphics[scale=0.22,clip]{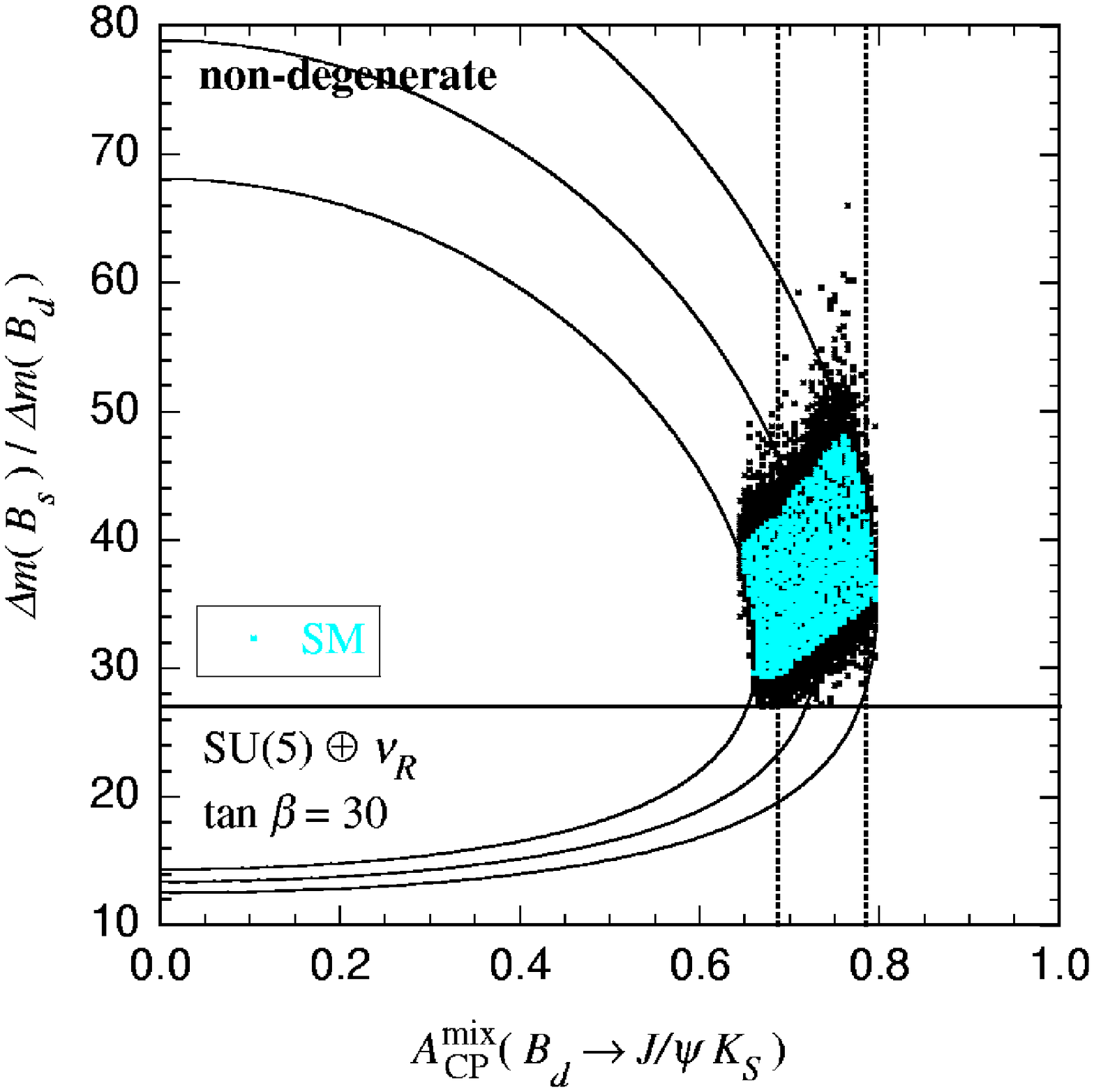}
\includegraphics[scale=0.22,clip]{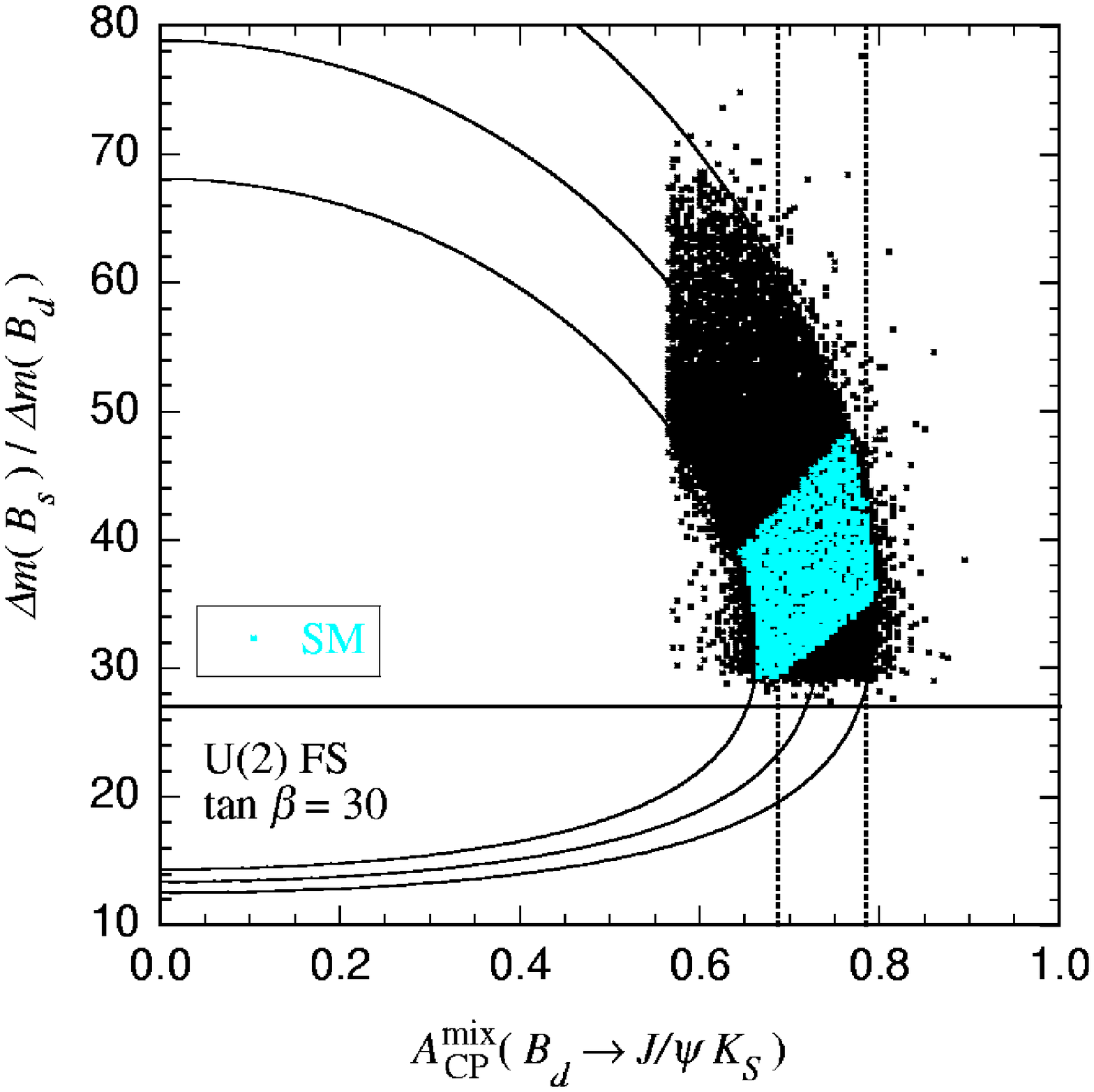}
\\
\includegraphics[scale=0.22,clip]{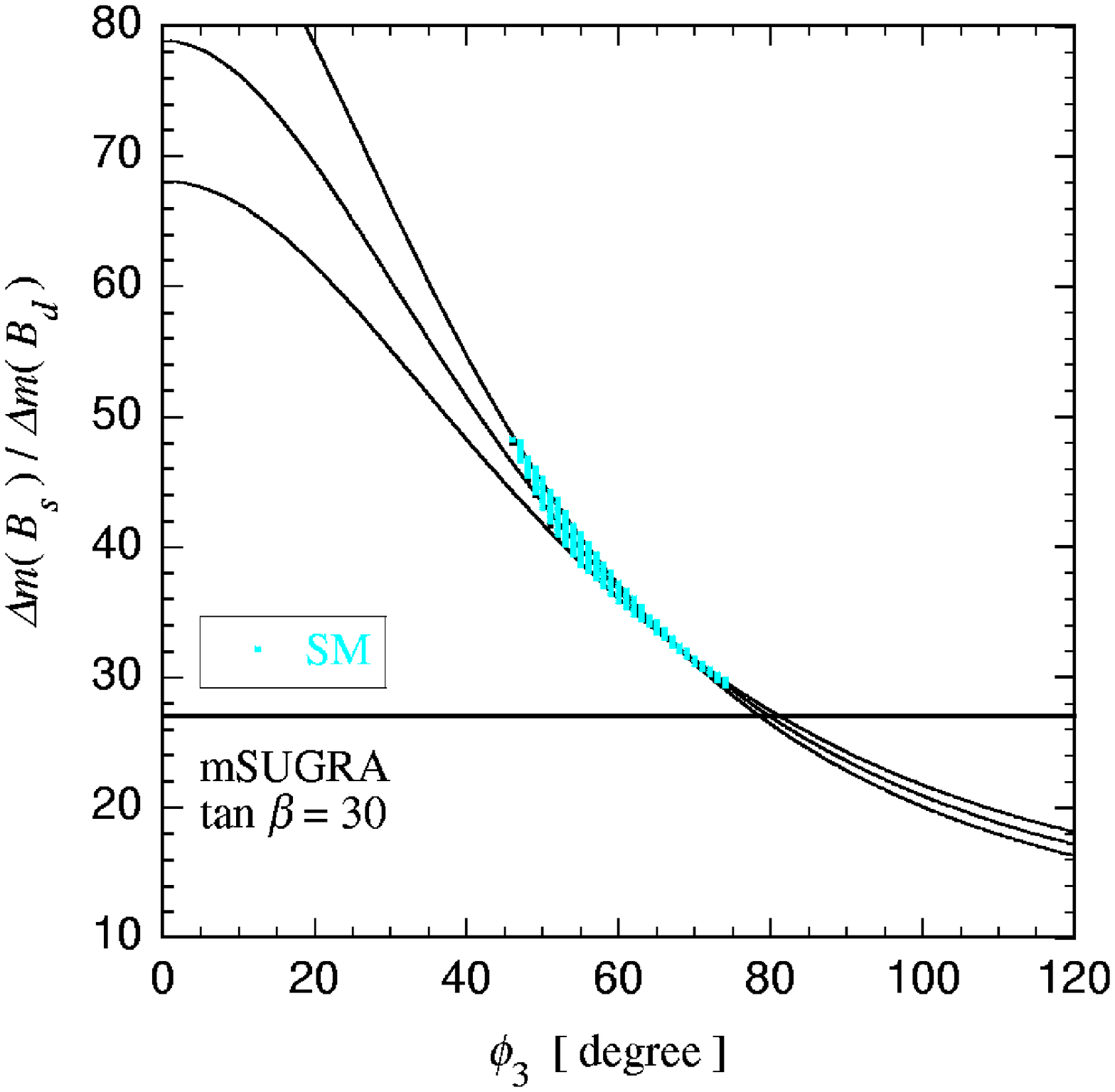}
\includegraphics[scale=0.22,clip]{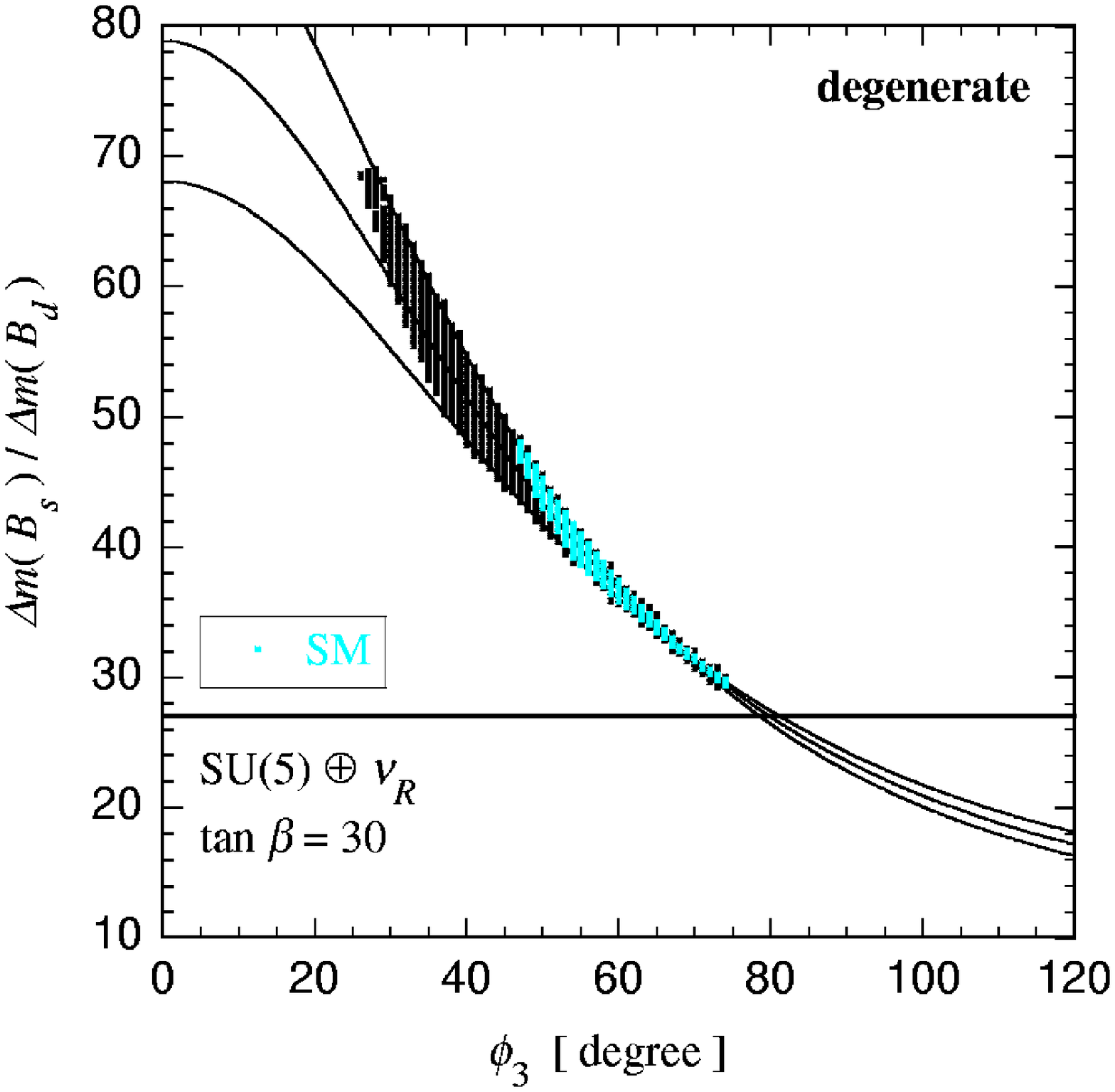}
\includegraphics[scale=0.22,clip]{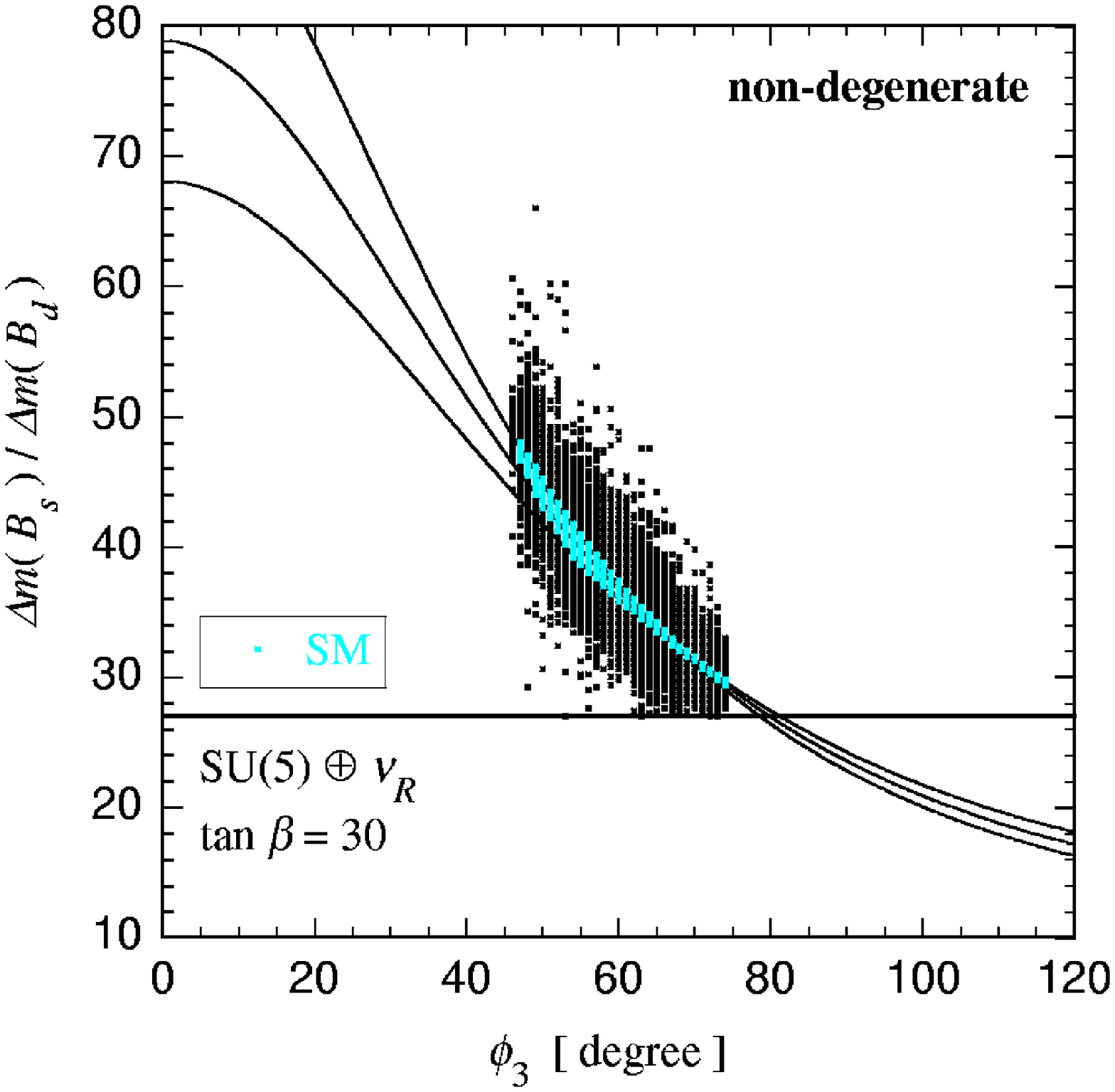}
\includegraphics[scale=0.22,clip]{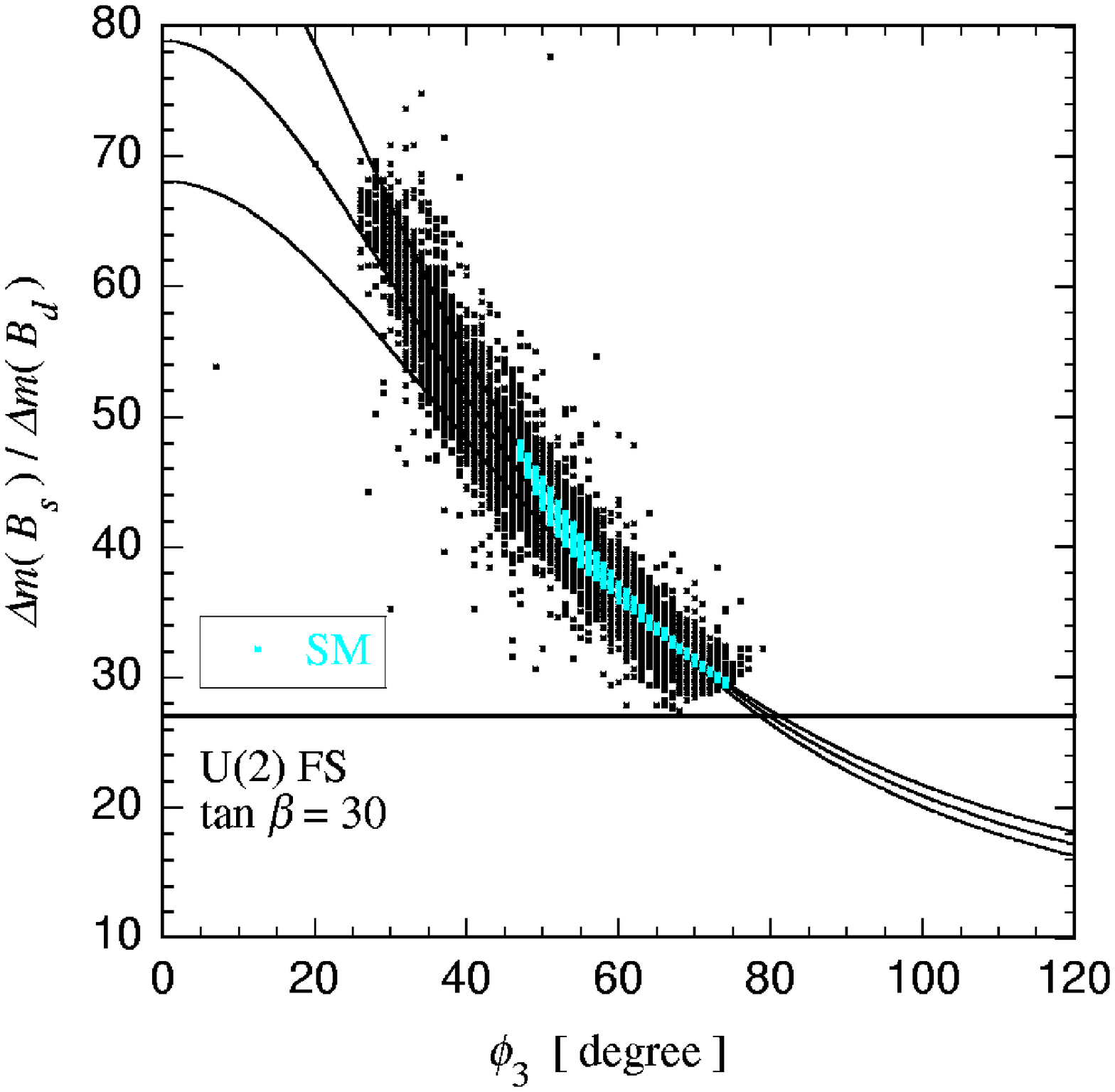}
\caption{
$\Delta m_{B_s}/\Delta m_{B_d}$ versus
$A_{\rm CP}^{\rm mix}(B_d\to J/\psi K_S)$ (upper row) and $\phi_3$
(lower row) in each model.
}
\label{fig:unitarity}
\end{figure}

In Fig.~\ref{fig:unitarity}, we show our numerical result
on the correlation among $\Delta m_{B_s}/\Delta m_{B_d}$, 
$A_{\rm CP}^{\rm mix}(B_d\to J/\psi K_S)$ and $\phi_3$.
Varying the free parameters in each model, we searched for allowed
regions consistent with the present experimental constraints.
In mSUGRA, possible deviations of all the three quantities are small.
In the degenerate case of the SU(5) SUSY-GUT with $\nu_R$, SUSY
contributions to the first and second generation mixing become large, so
that $\mu\to e \gamma$ and $\varepsilon_K$ are enhanced.
After imposing the experimental constraint on
${\rm B}(\mu\to e \gamma)$, sizable SUSY
contribution appears only in $\varepsilon_K$.
This correction of $\varepsilon_K$ leads to a change of the allowed
region of $\phi_3$ and eventually affects the possible region of 
$\Delta m_{B_s}/\Delta m_{B_d}$.
In the non-degenerate case, the neutrino Yukawa coupling matrix is
chosen such that $\mu\to e \gamma$ is suppressed
\cite{Ellis:2002fe}.
Then sizable contribution to $\Delta m_{B_s}$ is possible.
For the U(2) model, both patterns of deviation are possible.

\section{CP Asymmetries in $b\to s$ Decay}

The effective Lagrangian relevant for the $b\to s$ decays is
\begin{displaymath}
{\cal L}_{\rm eff} = C_{7L} {\cal O}_{7L} + C_{8L} {\cal O}_{8L}
+ (L\Leftrightarrow R) + (\mbox{four-quark}),
\end{displaymath}
where
${\cal O}_{7L} = \frac{e}{16\pi^2} m_b
 \bar{s} \sigma^{\mu\nu} b_R F_{\mu\nu}$,
${\cal O}_{8L} = \frac{g_3}{16\pi^2} m_b
 \bar{s} \sigma^{\mu\nu}T^a b_R G^a_{\mu\nu}$,
and the last term consists of the four-quark operators.
The SUSY contributions mainly affect the Wilson coefficients $C_{7L,R}$
and $C_{8L,R}$.
Each CP asymmetry depends on the coefficients differently.
$A_{\rm CP}^{\rm dir}(b\to s \gamma)$ is essentially determined by the
imaginary parts of $C_{7,8L}$, because the main contribution comes
from the interference between ${\cal O}_{7,8L}$ and the four-quark
operator $(\bar{s}\gamma^\mu c_L)(\bar{c}\gamma_\mu b_L)$.
$A_{\rm CP}^{\rm mix}(B_d\to K_{\rm CP} \gamma)$ is induced by the
interference of ${\cal O}_{7L}$ and ${\cal O}_{7R}$ so that 
$C_{7L} C_{7R}$ is relevant.
$A_{\rm CP}^{\rm mix}(B_d\to \phi K_S)$ is affected by $C_{8L,R}$.
Also, $|C_{7L}|^2+|C_{7R}|^2$ is constrained by the experimental value
of ${\rm B}(b\to s \gamma)$.

In the mSUGRA, SUSY contributions to $C_{7,8R}$ are very small.
The deviation in the ${\rm arg}\,C_{7L}$ is restricted by the EDM
experiments.
In the SU(5) SUSY GUT with $\nu_R$, the flavor mixing in the
$\tilde{d}_R$ sector is induced by the neutrino mixing and the GUT
interactions \cite{Moroi:2000tk}.
Thus the SUSY contribution to $C_{7R}$ can be large.
The constraints on ${\rm arg}\,C_{7L}$ is similar to the mSUGRA case.
The allowed region in the non-degenerate case is much larger than that
in the degenerate case since the $\mu\to e \gamma$ constraint is less
effective.
In the U(2) model,
large SUSY contributions are possible due to the flavor mixings and CP
phases in the squark sector.
SUSY contributions to $C_{8L,R}$ are similar to $C_{7L,R}$ in each
model.

\begin{figure}[htb]
\centering
\includegraphics[scale=0.45,clip]{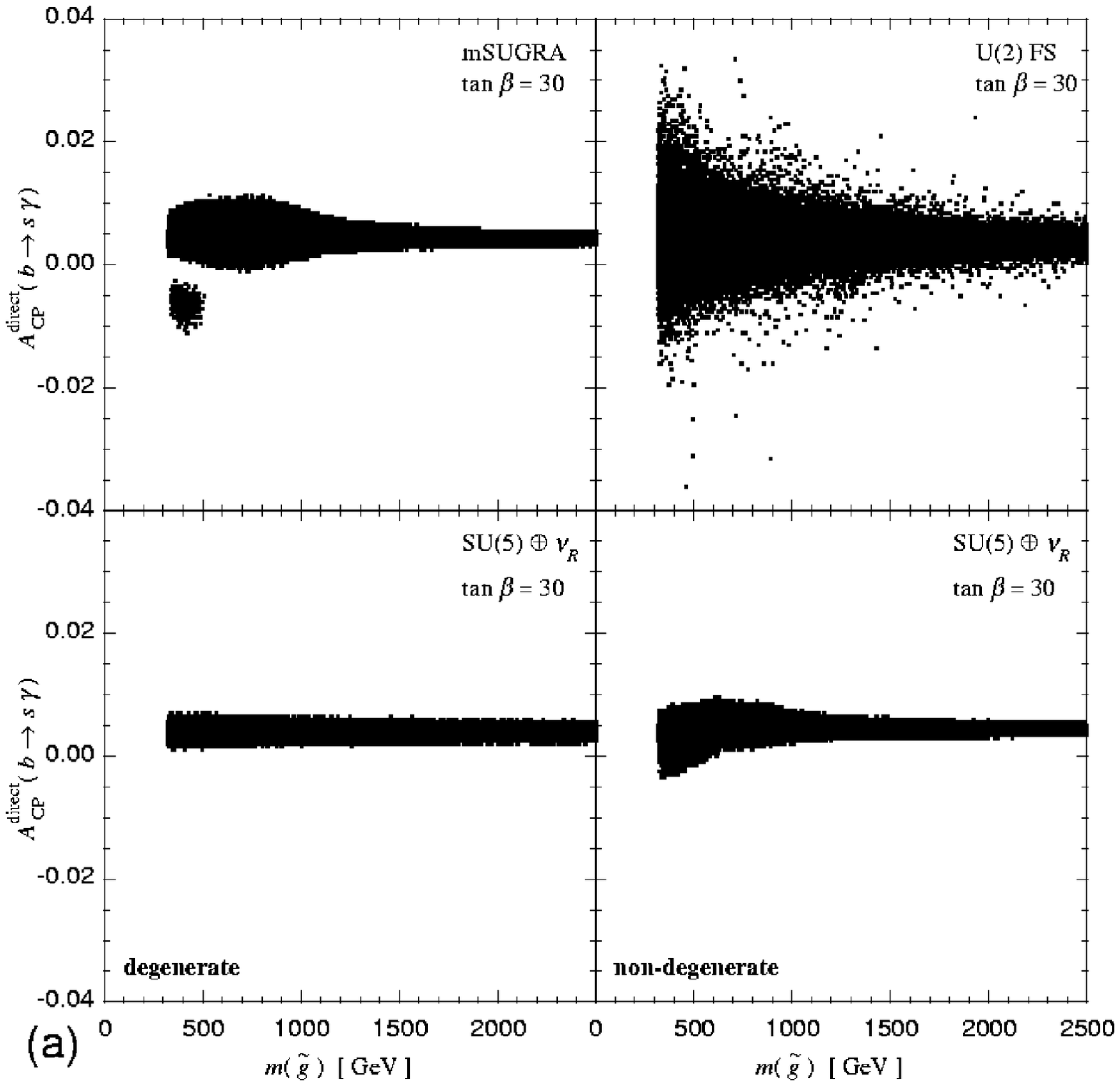}
\includegraphics[scale=0.45,clip]{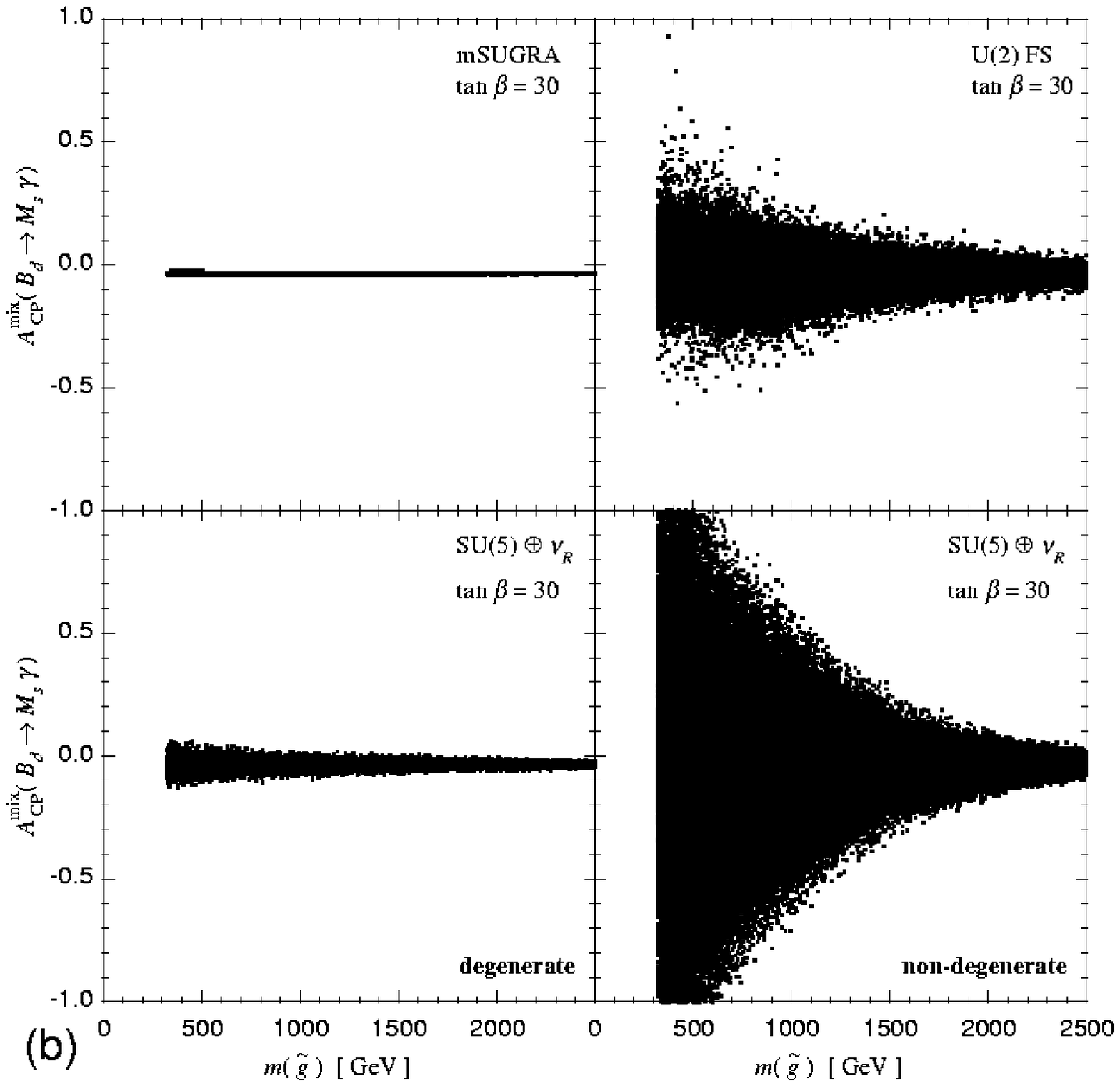}
\caption{
(a) $A_{\rm CP}^{\rm dir}(b\to s \gamma)$, and (b)
$A_{\rm CP}^{\rm mix}(B_d\to K_{\rm CP} \gamma)$
  as functions of the gluino mass.
}
\label{fig:bsgCP}
\end{figure}

In Fig.~\ref{fig:bsgCP}, we plot $A_{\rm CP}^{\rm dir}(b\to s \gamma)$ and
$A_{\rm CP}^{\rm mix}(B_d\to K_{\rm CP} \gamma)$ versus the gluino mass.
Possible value of $|A_{\rm CP}^{\rm dir}(b\to s \gamma)|$ is at most a
few percent compared with the SM prediction $\sim 0.5\%$.
The effect on $A_{\rm CP}^{\rm mix}(B_d\to K_{\rm CP} \gamma)$ is
significant in the SU(5) SUSY GUT with $\nu_R$ (non-degenerate case) and
in the U(2) model due to the SUSY contribution to $C_{7R}$.

\begin{figure}[htb]
\centering
\includegraphics[scale=0.6,clip]{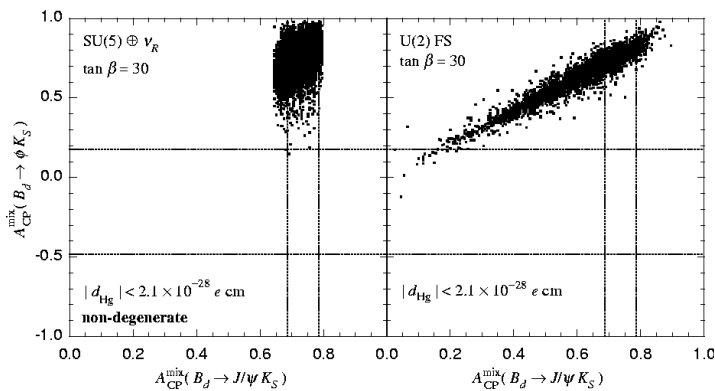}
\caption{
The correlation between 
$A_{\rm CP}^{\rm mix}(B_d\to \phi K_S)$ and 
$A_{\rm CP}^{\rm mix}(B_d\to J/\psi K_S)$.
}
\label{fig:phiKs}
\end{figure}

In Fig.~\ref{fig:phiKs}, we show the correlation between 
$A_{\rm CP}^{\rm mix}(B_d\to \phi K_S)$ and 
$A_{\rm CP}^{\rm mix}(B_d\to J/\psi K_S)$ for the non-degenerate case of
the SU(5) SUSY GUT with $\nu_R$ and the U(2) model.
In the SM, these asymmetries are the same.
The SUSY effect on the $B_d\to \phi K_S$ decay amplitude shows up as the
deviation from the equality.
In the non-degenerate case of the SU(5) SUSY GUT with $\nu_R$, $C_{8R}$
receives a large SUSY correction.
On the other hand, in the U(2) model, both $C_{8L}$ and $C_{8R}$ are
affected.
For the other two cases, the deviations from the SM relation are small.

\section{Conclusion}

In order to seek the possibility to distinguish SUSY models with $B$
physics, we studied the unitarity triangle analysis and the $b\to s$
decays.
In the unitarity triangle analysis, we showed that each SUSY model
provides the different pattern of the
correlation among $A_{\rm CP}^{\rm mix}(B_d\to J/\psi K_S)$,
$\Delta m_{B_s}$ and $\phi_3$, and could be distinguishable by precise
measurements.
We also explored SUSY effects on the CP asymmetries in
$b\to s \gamma$ and $B_d\to \phi K_S$ decays.
New signals in the mSUGRA and the degenerate case of the SU(5) SUSY
GUT with $\nu_R$ are relatively limited.
The non-degenerate case of the SUSY GUT exhibits large deviations
in $A_{\rm CP}^{\rm mix}(B_d\to K_{\rm CP} \gamma)$ and
$A_{\rm CP}^{\rm mix}(B_d\to \phi K_S)$.
The U(2) model predicts significant deviations in the $b\to s$ decays as
well as the unitarity triangle analysis.

Combining the above observation, we conclude that the study of the
unitarity triangle and $B$ decays could discriminate different SUSY
flavor models.
Since the flavor structure of the SUSY sector provides us a clue to the
origin of the SUSY breaking and interactions at high energy scales, $B$
physics will be essential to clarify the whole SUSY model.

\section{Acknowledgements}

The author would like to thank Y.~Okada, Y.~Shimizu, T.~Shindou and
M.~Tanaka.
This work is supported in part by the Grant-in-Aid for the 21st Century
COE ``Center for Diversity and Universality in Physics'' form the MEXT
of Japan.

\bibliographystyle{plain}

\end{document}